\documentclass[aps,prx,twocolumn]{revtex4-1}

\usepackage{graphics}
\usepackage{graphicx}
\usepackage{epstopdf}
\usepackage{amssymb}
\usepackage{amsmath}
\usepackage{hyperref}
\usepackage{latexsym}

\begin{document}

\renewcommand{\thefootnote}{\fnsymbol{footnote}} 
\renewcommand{\theequation}{\arabic{section}.\arabic{equation}}

\title{Effect of non-specific interactions on formation and stability of specific complexes}

\author{Dino Osmanovi\'c}
\author{Yitzhak Rabin}
\affiliation{Department of Physics, and Institute of Nanotechnology and Advanced Materials,
Bar-Ilan University, Ramat Gan 52900, Israel}

\date{\today}

\begin{abstract}
We introduce a simple model to describe the interplay between specific and non-specific interactions. We study the influence of various physical factors on the static and dynamic properties of the specific interactions of our model and show that contrary to intuitive expectations, non-specific interactions can assist in the formation of specific complexes and increase their stability. We then discuss the relevance of these results for biological systems.
\end{abstract}
\maketitle

\section{Introduction}
In physical and biological systems, there can exist different classes of interaction, termed \textit{specific} and \textit{non-specific} the former leading to the formation of specific complexes and the later referring to promiscuous interactions between any two molecules. In a biological context, specific interactions are critically important for the functioning of the cell and are closely associated with networks of protein-protein interactions (PPI)  \cite{Rual:2005,Deeds10012006,Levy:2008,Wodak:2013}.

Physically, an interesting question is how these proteins behave in the cellular environment, in terms of both their spatial distribution and dynamics. Since current analytical and computational methods are not powerful enough to capture the full complexity of a cell, one has to resort to simpler coarse-grained models. Deeds \textit{et al.}\cite{Deeds18092007} used a $3\times3\times3$ cube to represent a protein, where the cell was represented as a cubic lattice and equilibrium properties were studied using Monte Carlo simulations, whereas  McGuffee and Elcock\cite{McGuffee:2010} used Brownian dynamics to observe the behavior of tracer particle in a simulated cytoplasmic environment as a way of gauging how important the difference between \textit{in vitro} and \textit{in vivo} is. 

Of particular importance is the impact of the cellular environment on the formation of specific PPIs, and their stability. During its diffusion in the cell, any protein will feel many non-specific, promiscuous interactions with other proteins. The number of specific interactions a protein has may be swamped by this promiscuous environment. It has been estimated that the ratio between the typical energies of specific and non-specific attractions is around a factor of four\cite{MSB:MSB200848}. Therefore the impact that these non-specific interactions have on the cell is important in understanding the formation of and dynamics of specific PPIs. It is also an important signifier of the relevance of \textit{in vitro} experiments to the processes that take place in living cells: if the cellular environment  were to significantly change the properties of certain interactions, then this would imply that certain \textit{in vitro} experiments cannot be used to accurately infer \textit{in vivo} properties (for experimental comparison of the binding of some proteins under in vivo and in vitro conditions, see ref. \cite {Schreiber:2012}).

Since promiscuous interactions compete with specific ones, they are expected to have an adverse effect on protein-protein recognition and thus on the formation of specific protein complexes\cite{Janin:1996}.
However, not all non-specific interactions are necessarily harmful; for instance, in the question of how site-specific proteins can find a particular region of DNA\cite{Halford15052004} the role of non-specific interactions is crucial. By binding onto the DNA molecule, these proteins perform a one-dimensional instead of a three-dimensional search in order to find their targets efficiently \cite{Givaty:2009}. Similar mechanisms have also been proposed in other areas of biology e.g., in the context of selective transport of proteins through the nuclear pore complex\cite{TRA:TRA361}. Moreover, a possible way for expressing specific interactions in the cell involves first the formation of mostly non-specific encounter complexes \cite {Blundell:2006,Ubbink:2009}, which are then progressively rearranged to form specific bonds, and there is some experimental evidence to this effect \cite{Johansson:2014,Clore2007603}. Note that a similar mechanism has been implicated in many instances of protein folding which may begin with the formation of a molten globule driven by non-specific (hydrophobic) interactions, followed by rearrangement into the final native tertiary structure, see e.g., refs. \cite{Ptitsyn:1995,Pande:1998}. There is also an emerging literature based around the concept of phase separation of proteins in the cell, for which non-specific interactions are important \cite{Li:2012,OConnell:2012,Brangwynne:2013,Hyman:2012,Falkenberg:2013,Han:2012,Jacobs:2014}
In addition to these effects, there is also the effect of intracellular crowding that is dominated by non-specific steric repulsions that can affect both the equilibrium constants and the kinetics of protein complex formation\cite{Levy:2012}. 

In this paper, we employ a simplified model in order to study how non-specific interactions can affect specific association. By elucidating the mechanisms that control the formation and the stability of specific complexes, we hope to shed some light on familiar biological problems.

\section{Model}

In order to model both specific and non-specific interactions, we introduce an ensemble of $N$ spherical particles, interacting via a potential. We choose the Lennard-Jones potential as it contains both steric repulsion and long-range attractions necessary for a simple description of protein aggregation:
\begin{equation}
\phi(r_{ij})=4\epsilon_{ij}\left( \left(\frac{\sigma}{r_{ij}}\right)^{12} -\left(\frac{\sigma}{r_{ij}}\right)^{6}  \right)
\end{equation}
 We can also specify the potential to be purely repulsive, in which case:
\begin{equation}
\phi^{\mbox{rep}}(r_{ij}) = \begin{cases} 4\left( \left(\frac{\sigma}{r_{ij}}\right)^{12} -\left(\frac{\sigma}{r_{ij}}\right)^{6}  \right)+1& \mbox{if } r_{ij} < 2^{1/6} \\
0 &\mbox{otherwise} \end{cases}
\end{equation}
where $\sigma$ is the hard sphere diameter of the particle and $r_{ij}$ is the distance between particle $i$ and $j$. The $\epsilon_{ij}$ are elements of a symmetric $N\times N$ matrix specifying the interaction between particle $i$ and particle $j$. In our model, each of $N_S$ particles in the system has one (and only one) other particle with which it is assigned the specific interaction $\epsilon_{S}$, and with every other particle there is smaller non-specific interaction $\epsilon_{NS}$. In addition to this we can also introduce some other number $N_{I}$ of particles into the system which interact non-specifically with all $N-1$ other particles, where $N=N_I+N_S$. Note that when we are talking about ``specific" and ``non-specific" interaction, we assume that no covalent bonds are formed between the pairs and thus binding is reversible. Additionally, the manner in which the interactions are set up means that this model is fundamentally different from a two-component fluid, as each particle has only one other particle with which it has a strong specific interaction. 

We study this system using both Monte Carlo and molecular dynamics. 
In our system, there are several parameters of interest which we can vary such as the relative weight of the specific and non-specific interactions, $\epsilon_{\text{S}}$ and $\epsilon_{\text{NS}}$. 
We also vary the number of specific and non-specific particles present, $N_S$ and $N_I$ respectively, and the packing fraction $\eta$ (at fixed volume $V$), in order to explore the influence crowding has on the formation of specific complexes. We fix the temperature at $kT=1$, as biological systems do not typically vary in temperature too extremely. In our simulations we measure various statistical properties of the system relating to the properties of the specific bonds, such as the mean {\it complexation fraction} $\langle X_{S}\rangle$ defined as the ratio of the mean number of specific complexes to the maximal possible number of such complexes in the system, the average lifetime of a specific bond $\langle t_S\rangle$ and the average time between binding events (search time) $\langle t_F\rangle$. We define a complex as a specific pair that are within a distance of $1.6\sigma$ from each other.

\section{Results}

\begin{figure}[h!]
\begin{center}
\includegraphics[width=70mm]{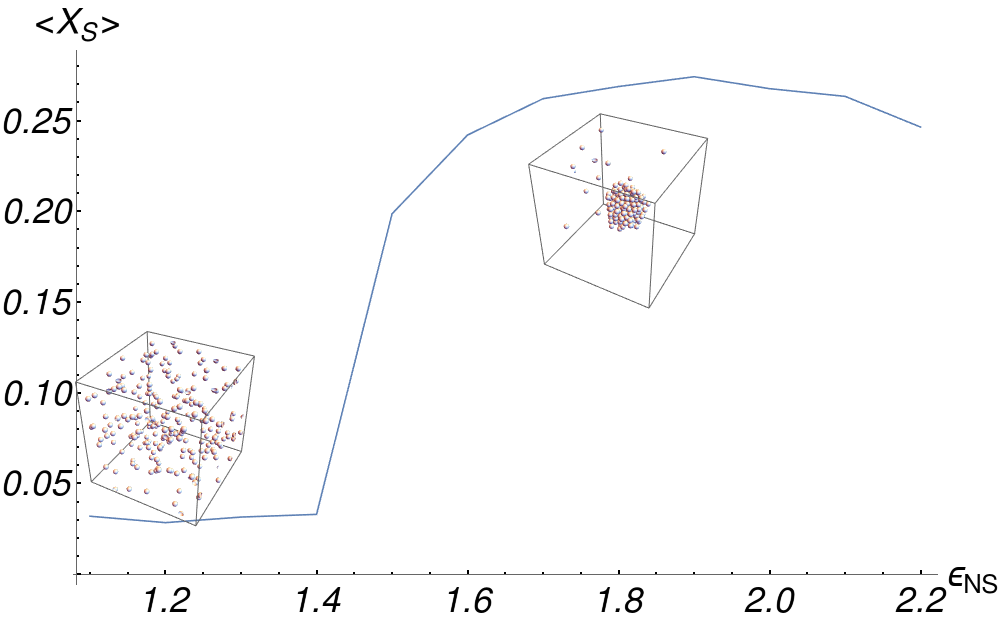}
\caption{The expressed number of specific bonds (with a total of 125 in the system) as a function of the non-specific interaction for $\epsilon_{S}=5$ and a packing fraction of 1\%. The insets show snapshots of the system at different values of $\epsilon_{NS}$  }
\label{fig:fig1} %
\end{center}
\end{figure}

Using Monte Carlo simulations, we begin by looking at the number of expressed specific bonds for $N_S=250$  and $N_I=0$ (the maximal possible number of specific bonds would then be 125) as we vary the strength of the non-specific interaction for a value of the specific interaction of $5$ and a packing fraction of 1\%.

As can be seen in fig. \ref{fig:fig1}, increasing the strength of the non-specific interaction at first has little effect on the number of specific complexes present in the system. When the strength of the non-specific interaction is further increased, there is a sharp rise in the complexation fraction, as the particles begin to agglomerate into a droplet that contains most of the particles in the system (gas to liquid transition). The physical mechanism behind the observed transition is clear: when the particles condense into a liquid droplet, their concentration increases, thus shifting the equilibrium constant of specific bond formation and the complexation ratio becomes much larger than when the particles move in a gas phase. At yet higher non-specific interaction strength, slow decrease of $\langle X_{S}\rangle$ is observed as non-specific interactions begin to compete with specific ones.
Increasing the other interaction parameter which we can control, the strength of the specific interaction $\epsilon_{S}$ (at fixed $\epsilon_{NS}$), increases the complexation ratio, as expected (not shown).


\begin{figure}[h!]
\begin{center}
\includegraphics[width=70mm]{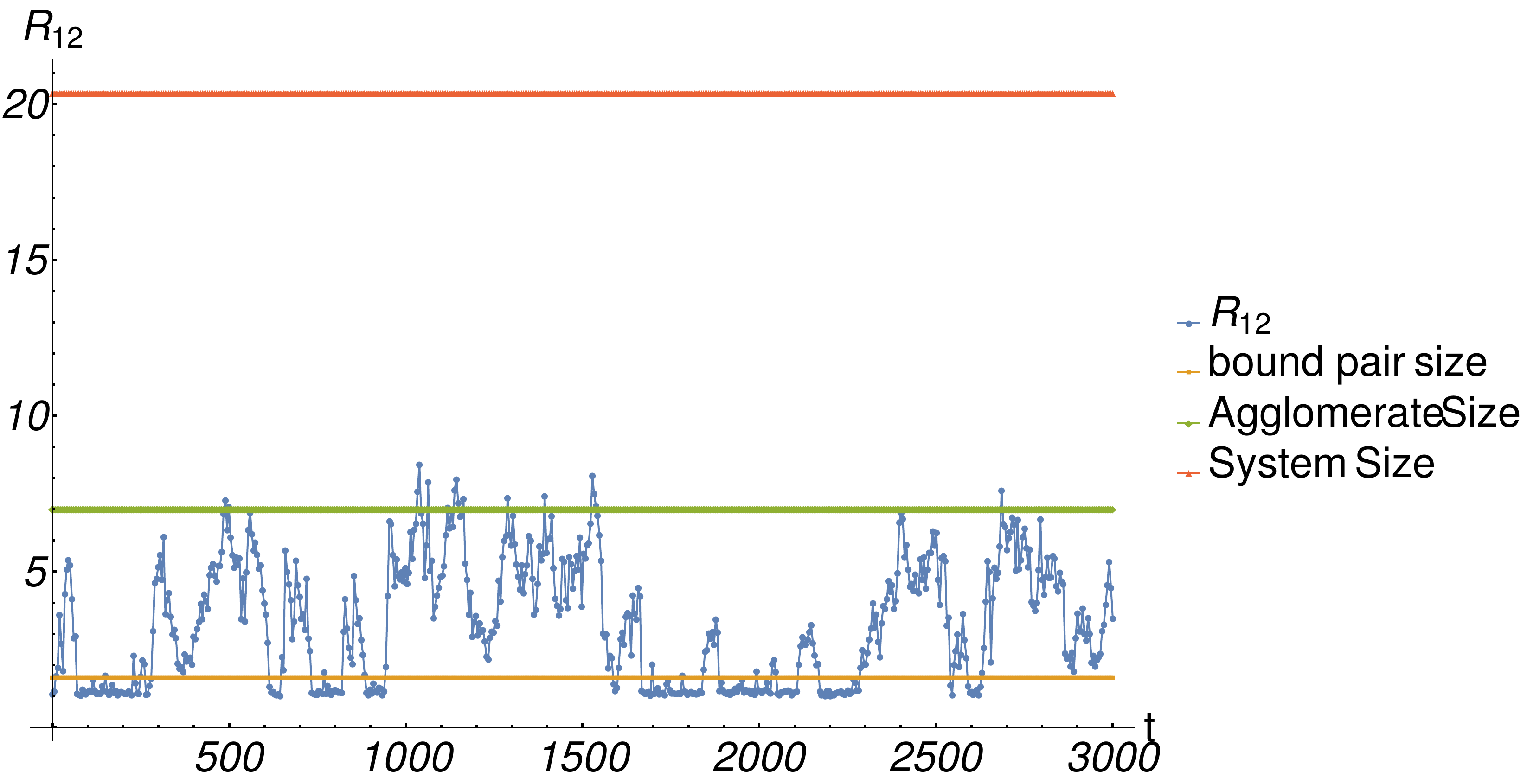}
\caption{The distance $R_{12}$ between a specific pair as a function of time. There are flat regions, where the pair is bound together, but thermal fluctuations can cause the pair to become separated. The parameters for this simulation are $N=N_S=250,\epsilon_{S}=5$, $\epsilon_{NS}=2.2$ and the packing fraction is 0.01.}
\label{fig:fig3} %
\end{center}
\end{figure}

\begin{figure*}
\begin{center}
\includegraphics[width=140mm]{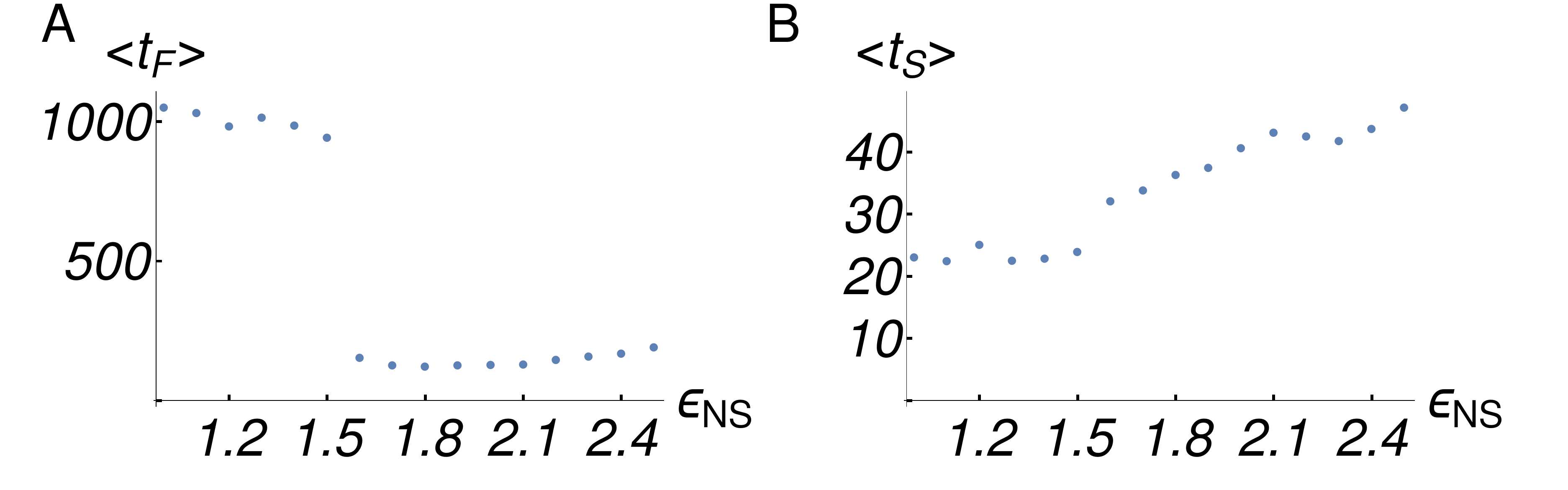}
\caption{A: The time in takes for specific pairs to find each other for a specific interaction of $\epsilon_{S}=5$ and $N=250$ at 1\% packing fraction as a function of the non-specific interaction. As the non-specific interaction is increased, the particles form an agglomerate, leading to shorter time necessary for pair ``search". B: How the lifetime of specific bonds depends on the strength of the non-specific interaction.  }
\label{fig:fig6}
\end{center}
\end{figure*}

 Using molecular dynamics (with periodic boundary conditions), we can also study the temporal dynamics of the system. For example, for a certain pair, we can look at how the distance between them varies in time, as can be seen in fig. \ref{fig:fig3}. For the set of interaction parameters in fig \ref{fig:fig3}, there exists a droplet (of diameter $\approx 7\sigma$) that contains nearly all of the particles the system.  Inspection of this figure shows that the particles in the pair can be bound together or separate but still remain in the droplet (events in which they detach from the droplet completely and move through the entire system, are not observed in the time window shown in fig \ref{fig:fig3}). A particular quantity of interest is the average lifetime of specific pairs $\langle t_S\rangle$. A large lifetime indicates that the pair are strongly bound.  Using the same temporal trajectories we can also measure the search time, i.e., the average time it takes for the specific pairs to find each other $\langle t_F \rangle$. These results are summarized in fig. \ref{fig:fig6} where $\langle t_S\rangle$ and $\langle t_F \rangle$ are plotted as a function of the non-specific bond strength, for the parameters in fig. \ref{fig:fig1}.

We find that whether the particle is in an agglomerate or not, has a large influence on both the stability of the bond as well as the time it takes for the specific pair to find one another. The agglomerated state stabilizes the bonds (by approximately a factor of 2) as well as greatly reducing the search time (by nearly an order of magnitude).

\begin{figure}
\begin{center}
\includegraphics[width=70mm]{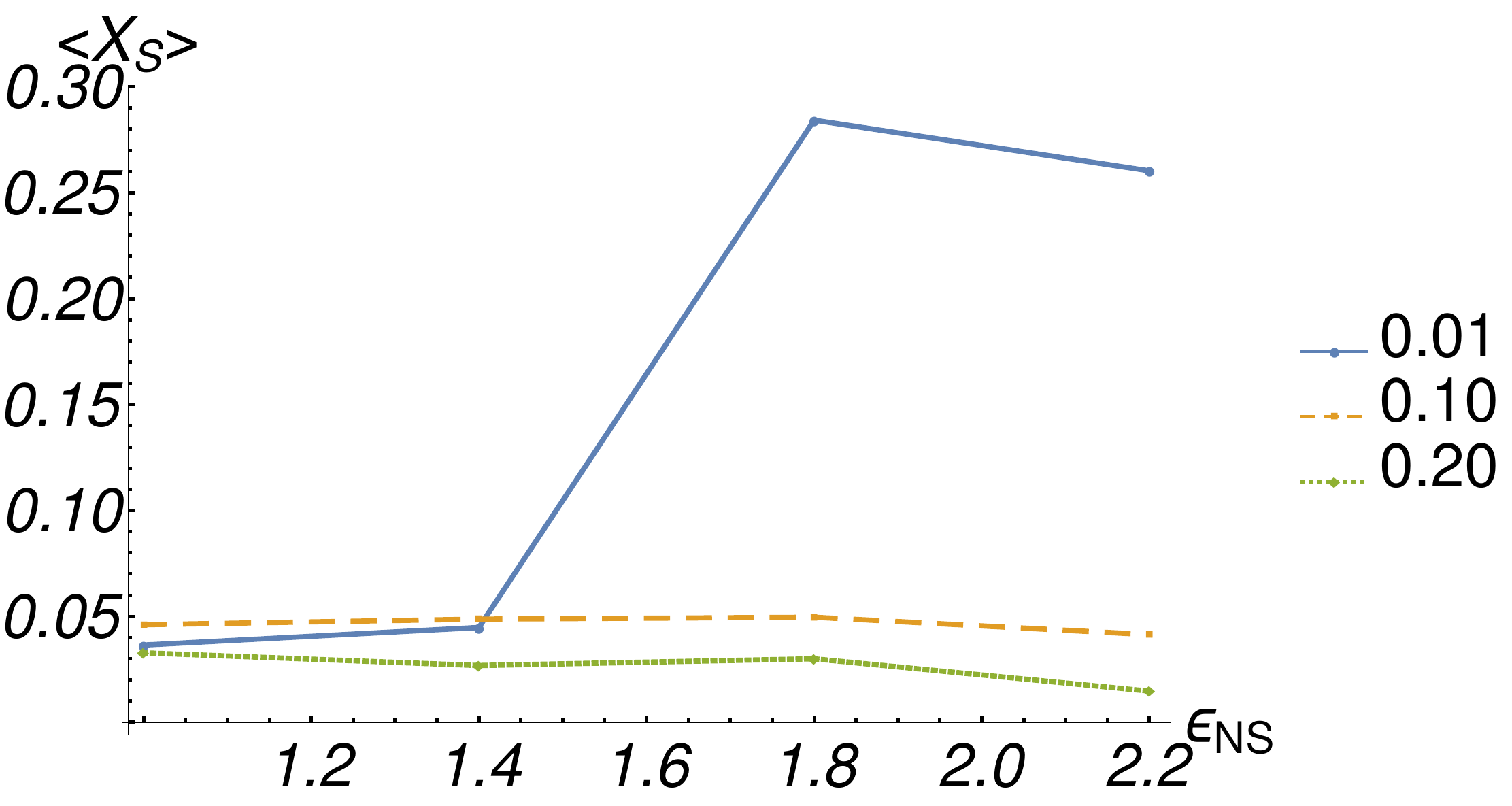}
\caption{Mean complexation fraction $\langle X_{S}\rangle$ against value of the non-specific interaction for different values of the packing fraction (0.01,0.1,0.2). Increasing the packing fraction leads to less importance of the non-specific interaction. }
\label{fig:fig9}
\end{center}
\end{figure}

So far, we have considered the situation where all the particles in the system have a specific partner. We now add other completely non-specific particles, $N_I$ (which do not have any specific partners and interact only via non-specific attraction), in order to see how crowding due to these non-specific particles affects  the complexation fraction of the specific pairs. The results are summarized in fig. \ref{fig:fig9}. We can see that as the packing fraction of non-specific particles increases, the sharp upward transition in the complexation fraction observed at $1\%$ packing fraction disappears and the expression rate becomes very low and  nearly independent of $\epsilon_{NS}$. This indicates that at higher packing fractions the increase of the concentration of the specific pairs due to agglomeration is too small to offset the increased competition with non-specific associations. Interestingly, in all cases increasing the packing fraction increases the mean lifetime of specific bonds (not shown).

\begin{figure}
\begin{center}
\includegraphics[width=80mm]{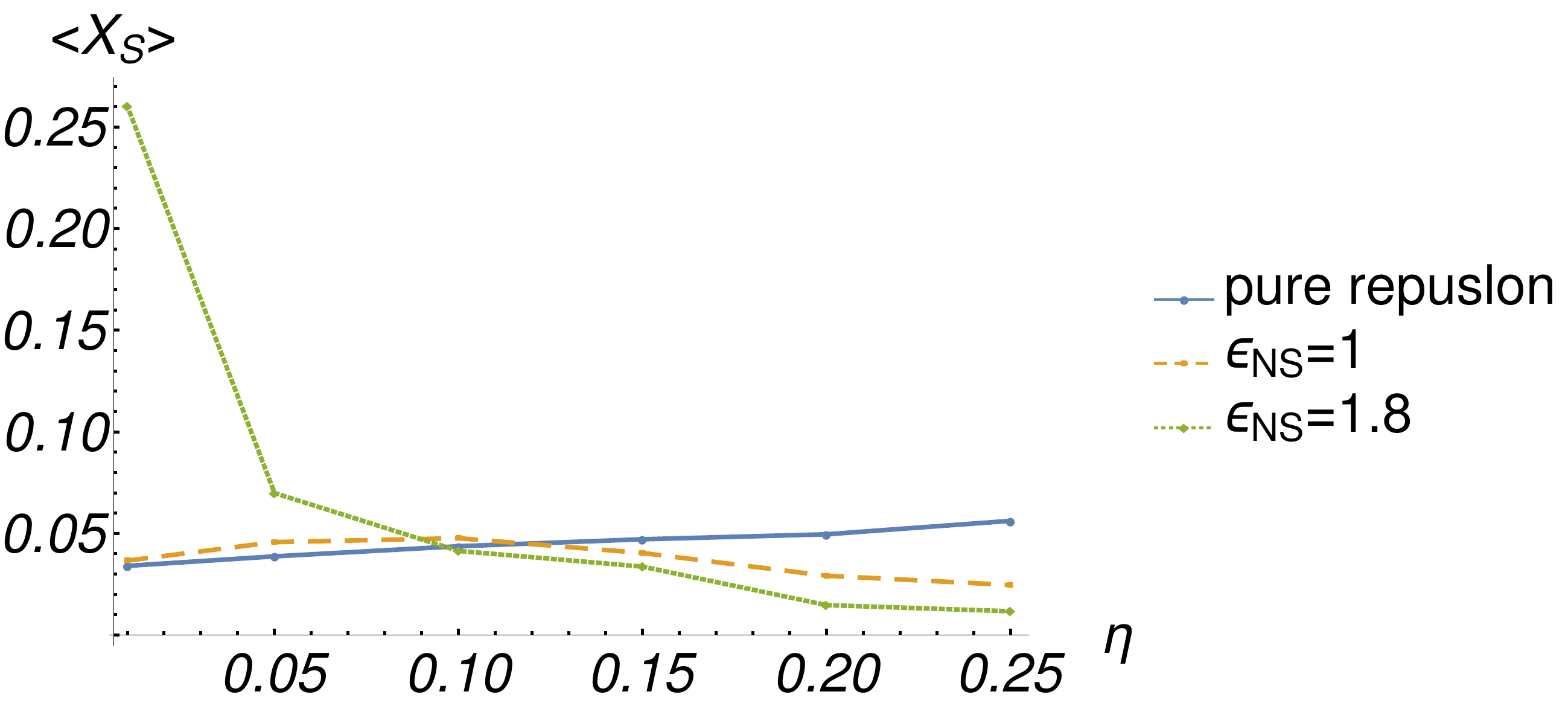}
\caption{The change in complexation as a function of packing fraction $\eta$ for purely repulsive non-specific interaction, $\epsilon_{NS}=1$ and $\epsilon_{NS}=1.8$ }
\label{fig:fig10}
\end{center}
\end{figure}

 If instead we plot how the complexation fraction varies with packing fraction for a chosen value of the non-specific interaction, the results are shown in fig. \ref{fig:fig10}. We find that the complexation fraction decreases monotonically with packing fraction for sufficiently attractive non-specific interactions and increases monotonically with packing fraction for repulsive non-specific interactions. For weakly attractive non-specific interactions there is a shallow maximum of the complexation fraction at packing fraction of about $10\%$. While strong non-specific attractions  are clearly more efficient in promoting the formation of specific pairs at low packing fractions (below $5\%$), the situation changes at higher packing fractions where repulsion-induced crowding becomes a more efficient enhancer of specific complexation, though the latter effect is much less pronounced than the former.

\begin{figure}
\begin{center}
\includegraphics[width=80mm]{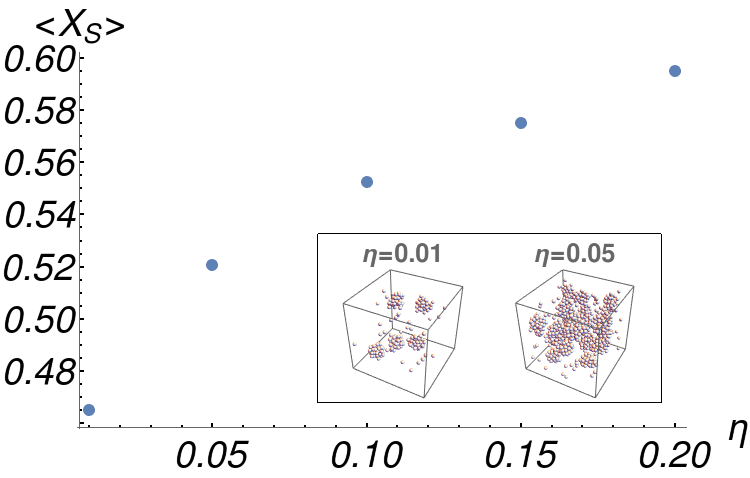}
\caption{The change in complexation as a function of packing fraction $\eta$ when there is a hierarchy of interactions with $\epsilon_S=5$ 
and  $\epsilon_W=2.2$ (see text).}
\label{fig:fig11}
\end{center}
\end{figure}

The reason for the adverse effect of increasing concentration on complex formation can be traced back to our particular choice of the interaction matrix $\epsilon_{ij}$ according to which all the non-specific interactions are identical. Thus, when we add more particles $N_I$ with with sufficiently large non-specific attraction $\epsilon_{NS}$, all the $N_S+N_I$ particles condense into a single large droplet in which the effect of increased concentration of specific pairs which promotes complex formation is opposed by the competition with non-specific associations which suppresses such specific complexes. Since the former effect decreases and the latter effect increases with the packing fraction (as the ratio $N_{I}/N_S$ increases), the agglomeration mechanism is no longer helpful at the rather high ($20-30\%$) packing fractions characteristic of proteins in cells. The  problem can be avoided and high complexation fractions can be achieved even at high concentrations if instead of a single large droplet, many smaller droplets are formed. This can be achieved by introducing a \textit{hierarchy} of specific interactions. 
For instance, consider the original $N_S=250,N_I=0$ system. Instead of having all the particles interact with each other non-specifically in the same way, we partition the system into sets of $50$ particles each, where in addition to strong ($\epsilon_{S}$) interaction between each specific pair of particles, all particles within each set weakly attract each other 
($\epsilon_{W}$) and interactions between particles not within the same set are purely repulsive. In this case, instead of forming a single aggregate of size $250$, there are several aggregates of $50$ particles each and the complexation fraction is significantly larger compared to the case where all the particles belong to the same set. As seen in fig. \ref{fig:fig11}, the complexation becomes even higher for this ``aggregates of $50$" case  as the packing fraction increases (increasing $N_S$ whilst $N_I$ remains 0), since this only increases the number of droplets while the size of each droplet remains unchanged (about $50$).

\section{Discussion and Conclusions}

We have presented a minimal model of a system with specific and non-specific interactions. While this model is too crude to capture the details of the complex interactions between proteins, there are some general features which will apply to any system with specific interactions.

\textbf{Non-specific interactions can help with the formation of specific pairs.} The naive expectation that specific and non-specific interactions always compete with each other isn't necessarily correct. Non-specific attractions can help in the formation of specific complexes provided they are strong enough to cause the particles to aggregate resulting in the formation of liquid droplets, as seen in fig. \ref{fig:fig1}. In such an agglomerated state, the complexation level of specific pairs increases, as the ``search space" is reduced compared to the case where the particles are moving freely throughout the entire system. There is some optimum range of values of the non-specific attractions which maximizes complexation by causing agglomeration. Further increasing the non-specific attraction beyond this point one reaches the ``competition" regime, which leads to decreasing complexation. One implication of such a mechanism is that both the search process and the stability of protein complexes can be assisted by non-specific attractions that result in non-uniform distribution of proteins in the cytoplasmic space or in membranes (the above mechanism applies equally well to the two-dimensional case). There are some indications that such clustering of proteins indeed takes place in live cells\cite{Narayanaswamy23062009}. The impact that phase transitions to liquid states can have on cellular function has been considered before\cite{Hyman:2012}. These transitions to liquid-like states are not merely confined to membranes, and the cell can have membran-eless compartments that perform functional biochemical reactions\cite{Brangwynne:2013}.For instance, several important biological molecules are found in liquid droplets inside the cell, such as RNA granules \cite{Han:2012}.  Our minimal model quantifies that this observed mechanism can increase complexation fraction by an order of magnitude. Similar principles can be applied to transient encounter complexes\cite{Tang:2006}, where non-specific associations between proteins take place before subsequent reorientation that results in specific bond formation.

\begin{figure*}[t!]
\begin{center}
\includegraphics[width=140mm]{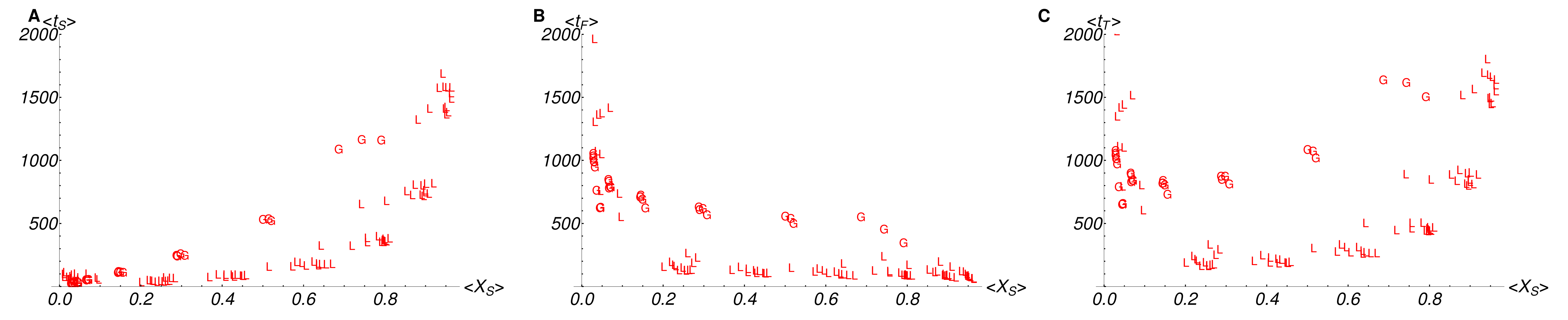}
\caption{Mean bond lifetime (A), search time (B) and turnover time ($\langle t_T\rangle=\langle t_S\rangle+\langle t_F\rangle$) (C) against complexation fraction.  The labels L or G refer to whether the state is agglomerated or non-agglomerated respectively. Increasing the complexation fraction generally leads to an increase in mean lifetime. As $\langle X_S\rangle$ approaches unity, the lifetime of the specific bonds becomes very large (A). Large specific lifetimes imply smaller search times (B), therefore an increase in complexation fraction leads to decreases in search times. The combination of these effects (C) gives the turnover time $\langle t_T\rangle$, which has a minimum around some complexation fraction. }
\label{fig:fig4}
\end{center}
\end{figure*}

\textbf{Large agglomerates are less helpful than small ones.} The mechanism outlined above is generally most effective for small numbers of particles. Small agglomerates have the largest complexation fraction, as can be seen from fig. \ref{fig:fig10}, when the particles are already in an aggregate (for $\epsilon_{NS}=1.8$); increasing the size of the agglomerate further by adding non-specific particles, $N_I$, leads to reduction in the complexation. This is essentially an entropic effect caused by the competition between specific and non-specific interactions in the aggregate. Larger aggregates will have more possible non-specific interactions present, shifting the system to decreasing complexation fraction as the size increases. The precise scaling of the complexation fraction with droplet size would depend on how the number of specific vs. non-specific interactions increases with increasing size of the droplet, but the fact that increasing the size of agglomerates would lead to decreased complexation would remain generally true so long as the number of non-specific interactions increases at a rate larger than that of specific interactions with increasing $N$. In most biological systems, large aggregates are associated with negative effects\cite{Ross:2004}, so the agglomeration mechanism discussed above would not be helpful if it involved very large aggregates. We demonstrated that this problem can be solved by introducing a hierarchy of interactions that lead to many small aggregates, as seen in fig. \ref{fig:fig11}. Most of the interactions with other components in the system have to be below some threshold so they don't form large aggregates, but stronger interactions within a certain small sub-class can  result in the formation of small clusters and increase the complexation fraction. This hierarchy of interactions leads to effective membrane-less compartmentalization in the form of droplets, as can be seen in certain cases in the cell \cite{Han:2012}. As has been discussed in previous literature \cite{Li:2012}, transitions to aggregated states lead to different spatial scales of organization, raising the possibility that even complicated environments like the cell can be organized on sub-micrometer length scales through effective membrane-less compartmentalization.

\textbf{The packing fraction affects complex formation.} There is much discussion in the biological literature of ``crowding"\cite{Konopka:2006,C3SM00013C}, and it is clear that the presence of crowding particles can affect the properties of the system. The effect depends on the precise nature of the interaction between the crowding particles and the complex-forming ones. As observed in fig. \ref{fig:fig10}, if the interaction is too attractive then adding these particles leads to increased competition and suppresses the formation of specific complexes. When the interactions with the crowding particles are purely repulsive, increasing the packing fraction leads to a gradual increase in the complexation as the addition of the other particles reduces the available volume to the specific pair and alters the free energy balance leading to greater complexation of specific pairs. In the intermediate region both these effects are present, leading to a shallow peak of complexation against packing fraction. In a biological context, it could be that the rather high packing fraction of proteins  in the cell ($20-30\%$) is actually helpful for the formation of specific complexes.

\textbf{Increasing the specific complexation fraction leads to exponentially increasing bond lifetimes.} We can ask the following: how do the mean lifetime and search time depend on the equilibrium complexation fraction of specific bonds $\langle X_S\rangle$? By combining all of our data for different conditions we examine the dependence of the bond lifetime $\langle t_S\rangle$, the search time $\langle t_F\rangle$ and the turnover time $\langle t_T\rangle=\langle t_S\rangle+\langle t_F\rangle$, on the complexation fraction $\langle X_S\rangle$. 
As can be seen in fig. \ref{fig:fig4}A, the mean lifetime of specific bonds increases monotonically with the specific complexation fraction and diverges as $\langle X_S\rangle\to 1$, but the behavior is different under conditions that correspond to the gas (G) and the liquid droplet (L) states; for the same $\langle X_S\rangle$, the lifetime is always higher in the gas phase (for the same$\langle X_S\rangle$ points in G correspond to higher values of $\epsilon_S$ than those in L). As shown in fig. \ref{fig:fig4}B, the search time decreases monotonically with the specific complexation fraction (it diverges as $\langle X_S\rangle\to 0$) but again the gas phase points lie above those that correspond to the liquid droplet state. The turnover time is plotted as a function of the specific complexation fraction in fig. \ref{fig:fig4}C; the two branches corresponding to the gas and the droplet phase exhibit  broad minima.  While we were not able to derive analytical expressions that describe these behaviors, a back-of-the-envelope estimate yields $\langle t_S\rangle/\langle t_F\rangle\approx \frac{\langle X_s\rangle}{1-\langle X_s\rangle}$, consistent with the observed divergence of $\langle t_S\rangle$ and $\langle t_F\rangle$ as $\langle X_s\rangle$ approaches unity and zero, respectively. Note that if the complexation fraction is 1 this implies that the binding is irreversible. Generally, if some parameter in the system (e.g., the strength of the specific bonds) is changed in order to increase the number of specific complexes, then the lifetime of the complexes increases in this manner with the complexation fraction. There are several consequences to this. If the bonds do not need to be reversible, i.e., after binding the functionality of the specific pair depends on them staying together all the time, then the strength of the specific interaction is essentially unbounded in terms of cost to functionality. However, if the binding needs to be reversible on some time scale then increasing complexation can lead to a cost to the system as the timescales of reversibility become larger than what is required. If the specific pair are bound together for too long but the components need to be recycled, either the system has to expend energy ( ATP) to separate them or this step would be a bottleneck in some dynamic process. Finally, there is an interesting possibility that the system may try to minimize the turnover time, by optimizing both the lifetime and the search time at some intermediate value of the complexation fraction.

We would like to comment on some features of our model that may appear to limit its applicability to biological systems, namely that each particle has one and only one partner with which it can form a specific complex and that all complexes contain only two particles each. Since a cell has numerous identical proteins that obviously have the same function and thus the same binding partners, and since many complexes consist of large numbers of proteins, these assumptions are clearly violated in the cellular environment. However, we believe that these limitations of our model do not change the our qualitative conclusions concerning the important role played by non-specific interactions in the formation of specific complexes. Another difficulty with application of our results to protein complexes is that the largest effects predicted by the simplest variant of our model correspond to particle concentrations that are about an order of magnitude lower than total protein packing fraction in cells. We have described one possible solution of this problem in terms of a hierarchy of specific interactions. Another possibility is that small droplets can be stabilized by surfactants (which may be other proteins; see e.g.,  ref. \cite{Rooney:1994}), analogously to mechanisms that stabilize micro-emulsions\cite{Langevin:1992}. Whether such mechanisms operate in cells is unknown at present and further  experimental studies are needed to determine the relevance of the various scenarios presented for biology.


\begin{acknowledgments}
Written correspondence with Emmanuel Levy and Naama Brenner is gratefully acknowledged. This work was supported by the I-CORE Program of
the Planning and Budgeting committee and the Israel Science Foundation grant 1902/12.
\end{acknowledgments}

\bibliography{paper}

\end{document}